\documentclass[%
reprint,
superscriptaddress,
amsmath,amssymb,
aps,
prl,
]{revtex4-1}

\usepackage{color}
\usepackage{graphicx}
\usepackage{dcolumn}
\usepackage{bm}

\begin {document}
\preprint{APS/123-QED}

\title{Dynamical Instability and Transport Coefficient in Deterministic Diffusion}

\author{Takuma Akimoto}
\email{akimoto@z8.keio.jp}
\affiliation{%
  Department of Mechanical Engineering, Keio University, Yokohama, 223-8522, Japan
}%

\date{\today}

\begin{abstract}
We construct both normal and anomalous deterministic biased diffusions to obtain the Einstein relation for their time-averaged
transport coefficients. We find that
the difference of the generalized Lyapunov exponent between biased and unbiased deterministic diffusions is 
related to the ensemble-averaged velocity. By Hopf's ergodic theorem, the ratios between the time-averaged velocity and the 
Lyapunov exponent for single trajectories converge to a universal constant, which is proportional to the strength of the bias.
We confirm this theory using numerical simulations.
\end{abstract}

\maketitle


{\it Introduction}.---Intrinsic randomness of time-averaged observables is of great interest in 
non-equilibrium statistical mechanics. Equilibrium systems are characterized by a few macroscopic 
observables, which are the time averages of microscopic observation functions. Macroscopic observables 
fluctuate around their ensemble averages in an equilibrium state. This means that the time averages of 
microscopic observation functions 
converge to their ensemble averages. In contrast, macroscopic observables such as the diffusion coefficient 
and fluorescence intensity cannot converge to a constant but show large fluctuations 
in non-equilibrium (non-steady) phenomena such as anomalous diffusions and intermittent phenomena \cite
{Golding2006, Szymanski2009,Brokmann2003}. In particular, diffusion coefficients of biological molecules in cells  \cite{Golding2006, Szymanski2009} and the fluorescence times in single nanocrystals \cite{Brokmann2003}
show large fluctuations, indicating ergodicity breaking.\par

Such intrinsic randomness of time-averaged observables is known to be universal in models characterized by 
power-law trapping-time 
distributions with divergence means \cite{Margolin2005,Margolin2006,Akimoto2008,He2008,Akimoto2010}. 
A typical example is a continuous time random walk (CTRW), which is a random walk with a random continuous trapping-time. 
In fact, CTRWs with infinite mean trapping-times
 show intrinsic randomness of the diffusion coefficient \cite{He2008, Miyaguchi2011}.
Although ergodicity, i.e., time average being equal to the ensemble average, does not hold in such systems, 
 time averages converge in distribution. This phenomenon is called
 {\it distributional ergodicity}  \cite{Miyaguchi2011a}D\par

In dynamical systems, distributional ergodicity is known to be the ergodicity in infinite measure systems because dynamical systems 
 relating to stochastic models with infinite mean trapping-times have {\it infinite invariant measures} \cite{Akimoto2010a}.
 In infinite measure systems, 
the time average for an observation function converges in distribution. The distribution of the time average is 
determined by properties of the observation function \cite{Akimoto2008}. 
In particular, the distribution of the time average of 
an $L^1(\mu)$ function $f(x)$ converges to the Mittag-Leffler distribution \cite{Aaronson1997}:
\begin{equation}
\frac{1}{a_n}\sum_{k=0}^{n-1} f \circ T^k \Rightarrow M_{\alpha},
\label{dist.lim.th}
\end{equation}
provided that $\int f d\mu \ne 0$, 
 where 
 $M_{\alpha}$ is a random variable with a Mittag-Leffler 
distribution of order $\alpha$, and $\Rightarrow$ denotes the convergence 
in distribution. The sequence $a_n$ is called the return sequence, which is relevant to non-stationarity. 
In deterministic subdiffusion, where the mean square displacement grows sublinearly $\langle x^2_n\rangle \propto n^{\alpha}$, 
the scaling of $a_n$ is the same as $a_n$ 
\cite{Akimoto2010}.\par

In diffusion processes, an external force $F$ generates a drift $V$. In general biased random walks including anomalous diffusion, 
the Einstein relation, 
\begin{equation}
V = \frac{FD}{2k_BT},
\end{equation} 
holds \cite{Barkai1998}, where $D$ is the diffusion coefficient under no bias, $k_B$ is the Boltzmann constant,
 and $T$ is the temperature. 
Diffusion properties in hyperbolic chaotic dynamical systems are well known. Using the escape rate formula, one can show that 
the diffusion coefficient is equal to the difference between the Lyapunov exponent and Kolmogorov-Sinai entropy
 \cite{Gaspard1990}. Moreover, in a Lorentz gas, 
the largest Lyapunov exponent is suppressed by an external field and there is a relation 
between the Lyapunov exponents and a bias \cite{Dettmann1995, Morriss1996}. 
However, little is known about the relation between dynamical instability and the transport property under a bias 
. \par

In this Letter, we derive a dynamical system for arbitrary biased and unbiased CTRWs. 
Furthermore, we explain the linear response to a small bias and the Einstein relation in general 
deterministic diffusions. Using the Lyapunov exponent, we obtain a relation between the macroscopic transport coefficient 
and the microscopic chaos in a deterministic subdiffusion.
In particular, we relate the ensemble-averaged velocity to
 the difference of dynamical instabilities between an unbiased and a biased dynamical system.
Using Hopf's ratio ergodic theorem, we show that the ratio between the time-averaged velocity and the Lyapunov exponent 
converges to a universal constant. Moreover, we find that the universal constant is proportional to a bias, and the proportional 
constant is determined by the diffusion coefficient and the Lyapunov exponent under no bias. \par

{\it Deterministic biased diffusion model}.---Dynamical systems exhibiting diffusion are represented by 
 one-dimensional map $T(x)$ with translational symmetry, $T(x+L)=T(x) +L$ $(L=0, \pm 1, \ldots)$ 
\cite{Geisel1984}. The map $T(x)$ can be reduced to a map on unit interval $[-1,2, 1/2]$, $R(x)$. 
Here we construct a dynamical system corresponding to the CTRW. The CTRW is defined 
by a jump length distribution $l(x)$ and a waiting time distribution $\psi (t)$. For simplicity, a jump length distribution 
is set to $l(x)=p\delta(x-1) + q\delta(x+1)$ $(p+q=1)$. Since waiting times in CTRWs are independently and identically distributed 
random variables, we can construct a one-dimensional piecewise linear intermittent map $R(x)$ with 
the same waiting time distribution $\psi(t)$ 
\cite{Akimoto2010b}. The probability for the right jump $p$ is represented by the length of the interval $[c,1/2]$, 
which is mapped to the right side in a neighboring cell. As shown in Fig.~1, the reduced map on $[1/4,1/2]$ is given by an asymmetric   
 piecewise linear map. We note that
the scaling exponents of the waiting time distributions for the right and left jumps are the same. However,
the waiting time distribution for the interval $[0,1/4]$ are slightly different from that for the interval $[-1/4,0]$, 
since the derivatives $R(x)$ at $x=c+0$ and $c-0$ are different in biased models \footnote{Strictly speaking, the 
waiting time distribution is different from each biased dynamical system. However, the tail of the waiting time distribution 
is exactly the same. Therefore, this difference does not affect a bias but affect the mean of the total number of jumps 
$\langle \Delta n(k) \rangle$.}. 
When the mean waiting time diverges, $\psi(t) \propto t^{-1-\alpha}$ $(0<\alpha<1)$, $R(x)$ has an infinite invariant measure 
\cite{Akimoto2010a} and the map $T(x)$ causes subdiffusion, i.e., $\langle x_n^2 \rangle \propto n^{\alpha}$ 
 \cite{Akimoto2010}.\par
 
\begin{figure}
\includegraphics[height=.55\linewidth, angle=0]{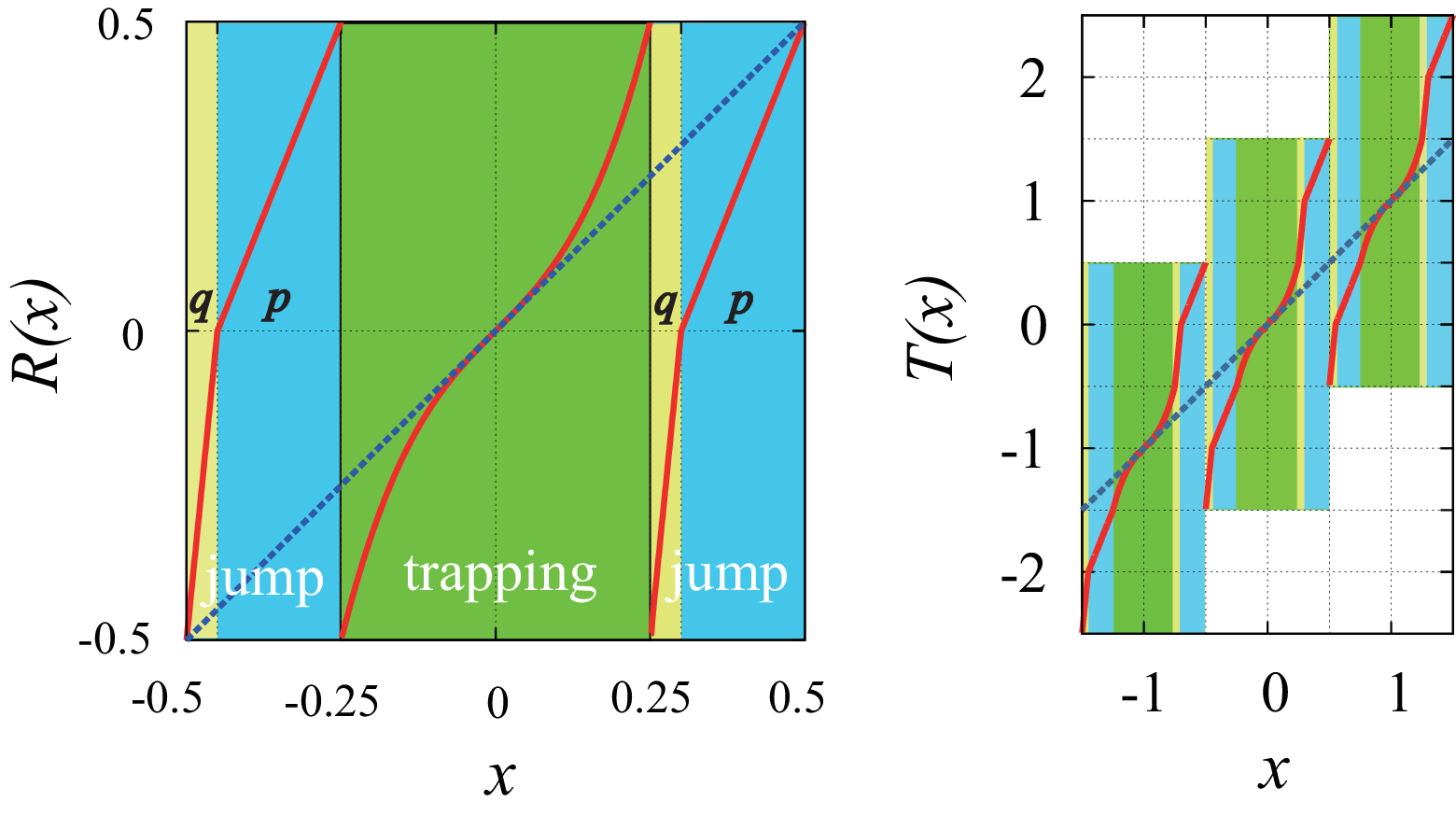}
\caption{ (Color online) Reduced map $R(x)$ and the corresponding original map $T(x)$ 
[$z=3.0$ and $c=0.3$ in Eq. (\ref{e.reduced_map})].}
\end{figure}

{\it Einstein relation}.---The Einstein relation refers to the relation between the velocity under a bias and 
the diffusion coefficient under no bias.
 We obtain the Einstein relation for the time-averaged transport coefficients in normal and anomalous deterministic diffusions.
The time-averaged mean square displacement (TAMSD) and the time-averaged drift (TAD) are defined by
\begin{equation}
    \overline{\delta x^2_{m}}(N;x_0)
    \equiv
    \frac{1}{N}\sum_{k=0}^{N-1} (x_{k+m} -x_k)^2
\end{equation}
and
\begin{equation}
\overline{\delta x_m}(N;x_0) = \frac{1}{N} \sum_{k=0}^{N-1}(x_{k+m}-x_k),
\end{equation}
respectively. Unlike the ensemble-averaged mean square displacement, TAMSD shows normal diffusion even when $\alpha<1$ 
\cite{Akimoto2010}. Therefore, the time-averaged diffusion coefficient is defined by $\overline{D} \equiv 
\overline{\delta x^2_{m}}(N;x_0)/m$. In finite measure cases, the time-averaged diffusion coefficient is equal to the 
ensemble-averaged one. 
The ensemble-averaged drift $\langle \delta x_m \rangle_F$ is given by $\langle \delta x_m \rangle_F \equiv \langle x_{m}-x_0 \rangle_F$, where $\langle \cdot \rangle_F$ refers to the ensemble average under a bias $(c\ne 0.375)$. 
Let $N_m$ be the total number of jumps until time $m$,
 then we have  $\langle \delta x_m \rangle_F=(p-q)\langle N_m \rangle_F$. 
 Here we assume that the injection to the set $[-1/2,-1/4)\cup (1/4,1/2]$
 is uniform (Assumption A). This assumption is exact when the map is a piecewise linear map. It follows that $p=2-4c$ and $q=4c-1$. 
In normal diffusion, $\langle N_m \rangle_F$ is given by $\mu(J^c) m$, where $\mu$ is an invariant measure of $R(x)$ and 
$J=[-1/4,1/4]$. 
In subdiffusion, $\langle N_m \rangle_F \propto m^{\alpha}$. Then, we have 
\begin{equation}
\langle \overline{\delta x_m}(N) \rangle_F = (p-q) \sum_{k=0}^{N-1} \frac{\langle N_{k+m}\rangle -
\langle N_k \rangle}{N} \cong \langle \overline{V}\rangle_F m,
\end{equation}
where we define the time-averaged velocity by $\overline{V}\equiv \overline{\delta x_m}(N)/m$.\par

Let $p=W_0e^{-F/2k_BT}, q=W_0e^{+F/2k_BT}$, and 
$ \Delta n(k)$ be the total number of jumps in the interval $[k,k+m]$, so that 
$\langle x_{k+m}-x_k \rangle_F \cong (p-q) \langle \Delta n(k)\rangle_F$.
Here we assume $\langle \Delta n(k)\rangle_F=K\langle \Delta n(k)\rangle$ (Assumption B). 
$K$ does not depend on $c$  for a small bias
 because the scaling of $\langle \Delta n(k)\rangle_F$ is the same
for all $c$. It follows that 
$\langle x_{k+m}-x_k \rangle_F=K(p-q) \langle \Delta n(k)\rangle \cong K(p-q) \langle (x_{k+m}-x_k)^2\rangle$.
We have 
\begin{equation}
\langle \overline{V} \rangle_F  = K \langle \overline{D} \rangle \varepsilon,
\label{linear.response}
\end{equation}
where $\varepsilon = 3-8c$.
Thus, we obtain the Einstein relation,
\begin{equation}
\langle \overline{V} \rangle_F  = \frac{\langle \overline{D} \rangle F'}{2k_BT},\quad F\rightarrow 0,
\label{einstein.ensemble}
\end{equation}
where $F'=KF$, which is not equal to $F$. It seems that the Einstein relation is violated. However, 
 the change in the waiting time distribution due to the bias results in the violation of the Einstein relation.
 Therefore, the Einstein relation holds under 
the bias $F'$, because the difference between the waiting time distributions for the right and the left does not generate 
a drift. 
We note that the ensemble average in the Einstein relation is essential in anomalous subdiffusion because the time-averaged 
transport coefficients $\overline{D}$ and $\overline{V}$ are intrinsically random. 
When the time-averaged velocity $ \overline{V}$
and the time-averaged diffusion coefficient $\overline{D}$ are measured independently, $\overline{V}/\overline{D}$ becomes 
random, namely, the effective temperature, $F\overline{D}/2k_B\overline{V}$, becomes random \cite{Shemer2009}.\par

{\it Relation between the Lyapunov exponent and the velocity for single trajectories}.---The Lyapunov exponent 
is defined by
\begin{equation}
\overline{\lambda} (N,x_0) \equiv \frac{1}{N} \sum_{k=0}^{N-1} \ln |T'(x_k)|.
\end{equation}
By Hopf's ergodic theorem \cite{Aaronson1997}, 
\begin{equation}
\frac{\sum_{k=0}^{n-1} f_m(T^k x)} {\sum_{k=0}^{n-1} g( T^kx)}  \rightarrow C_m=
\frac{\int_0^{1/2} f_m d\mu }{\int_0^{1/2} g d\mu}
\label{hopf}
\end{equation}
holds for almost all initial points of $x$, where $f_m(x)=T^m(x)-x$ and $g(x)=\ln |T'(x)|$.
 Therefore, the time-averaged velocity $\overline{V}(N;x)\equiv \overline{\delta x_m (N,x)}/m$ and the 
 Lyapunov exponent $\overline{\lambda}(N;x)$  satisfy the following relation, 
 $\overline{V}(N;x)/\overline{\lambda}(N;x) \rightarrow C_m/m$ as $N\rightarrow \infty$ 
for almost all $x$. 
Since $\overline{\delta x_{m}}(N) \propto m$ for a large $m$, $C_m/m$ does not depend on $m$. 
Let $C_m/m$ equal $\chi_V$, then we can obtain the relation between the velocity and the Lyapunov 
exponent for almost all initial points of $x$:
\begin{equation}
\frac{\overline{V}(N;x)}{\overline{\lambda}(N;x)} \rightarrow \chi_V \quad {\rm as}~N\rightarrow \infty.
\end{equation}
In general, the ensemble average of the Lyapunov exponent decreases with increase in 
the ensemble-averaged velocity (see Fig.~3).  
Nevertheless, the time-averaged velocity is proportional to the Lyapunov exponent. 
The constant $\chi_V$ is a universal constant, which does not depend on an initial point. Using the ensemble average 
and  Eq.~(\ref{linear.response}), we obtain 
\begin{equation}
\chi_V = \frac{\langle \overline{V}\rangle_F}{\langle \overline{\lambda}\rangle_F}=
\frac{K\langle \overline{D}\rangle}{\langle \overline{\lambda}\rangle_F} \varepsilon.
\end{equation}
\par

{\it Difference of the Lyapunov exponent}.---We consider the difference of the generalized Lyapunov 
exponent between an unbiased and a biased dynamical system. The generalized Lyapunov exponent is the ensemble average 
of the normalized Lyapunov exponent \cite{Korabel2009,Akimoto2010a}
\begin{equation}
 \lambda  \equiv \left\langle \lim_{n\rightarrow \infty} \frac{1}{a_n} \sum_{k=0}^{n-1} g(x_k)
\right\rangle,
\end{equation}
where $\langle \cdot \rangle$ represents the ensemble average of the initial points. Note that the density
 is absolutely continuous with 
respect to the Lebesgue measure and satisfies the condition $\langle g(x_0) \rangle < \infty$.
The assumption B, $\langle \Delta n(k)\rangle_F=K\langle \Delta n(k)\rangle$, means that the statistical property of 
the reinjection to the indifferent fixed points is almost the same. That is, the generalized Lyapunov exponent restricted 
to $J=[-1/4,1/4]$ is almost the same. 
The difference of the generalized Lyapunov exponent, $\Delta \lambda$, is defined by
$
\Delta \lambda = \langle \overline{\lambda}(x) \rangle - \langle \overline{\lambda}(x) \rangle_F.
$

By assumption A and B,
\begin{equation}
\Delta \lambda = \frac{1}{a_n}\sum_{k=0}^{n-1} (\langle g(x_k)1_{J_c}(x_k)\rangle - \langle g_c(x_k)1_{J_c}(x)\rangle_F).
\end{equation}
Thus, 
\begin{equation}
\Delta \lambda = \int_{J_c} g(x)d\mu - \int_{J_c}g_c(x)d\mu
 = \Delta \lambda_{J^c} \mu (J^c),
\end{equation}
where $\Delta \lambda_{J^c}$ is the difference of the generalized Lyapunov exponent restricted to $J^c$. Since 
the injection to $J^c$ is uniform (Assumption A), we have 
$
\Delta \lambda_{J^c} = \ln 2 + p\ln p + (1-p) \ln (1-p).
$ 
The generalized velocity and its maximum are defined as $V \equiv \lim_{n\rightarrow\infty} \langle \delta x_n\rangle_F/n^{\alpha}$ and 
 $V_{\max} \equiv \lim_{n\rightarrow\infty} \langle N_n\rangle_F/n^{\alpha}$, respectively. We use
$
\frac{V}{V_{\max}} = p-q.
$
Then, the probability $p$ 
is written by $V$ and $V_{\max}$: $p=\left(1 + V/V_{\max} \right)/2.$
Using $V$ and $V_{\max}$, we obtain the relation between the difference of the generalized Lyapunov exponent and the generalized 
velocity:
\begin{equation}
\Delta \lambda = 
\frac{\mu (J^c)}{2} S\left(\frac{V}{V_{\max}}\right),
\label{theoretical.curve}
\end{equation}
where $S(x) = (1+x)\ln (1+x) +(1-x) \ln (1-x)$. Therefore, the constant $\chi_V$ is given by
\begin{equation}
\chi_V = 
\frac{K\langle \overline{D}\rangle}{\langle \overline{\lambda}\rangle} \varepsilon + O(\varepsilon^2).
\label{chiV}
\end{equation}
Note that the proportional constant is determined by  
the diffusion coefficient and the Lyapunov exponent under no bias.

{\it Example for deterministic subdiffusion}.---We demonstrate numerical results for deterministic subdiffusion. 
A piecewise linear map is a good approximation for an intermittent map. 
We consider the following intermittent reduced map:
\begin{equation}
 \label{e.reduced_map}
 R(x)=\left\{
 \begin{array}{ll}
 \frac{x-3/4+c}{2c-1/2}  & x\in [-\frac{1}{2}, -\frac{1}{4}+c),\\
 \frac{x-3/4+c}{1-2c}  & x\in [-\frac{3}{4}+c, -\frac{1}{4}),\\
 x-4^{z-1}(-x)^z & x\in [-\frac{1}{4},0),\\
 x+4^{z-1}x^z\quad  & x\in [0,\frac{1}{4}),\\
 \frac{x-c}{2c-1/2}  & x\in [\frac{1}{4}, c),\\
 \frac{x-c}{1-2c} & x\in [c, 1/2].
 \end{array}
 \right.
\end{equation} 
The invariant density of the reduced map is given by $d\mu = h(x)|x|^{1-z}dx$, which means an infinite invariant measure 
for $z\geq 2$ \cite{Thaler1983}. Since the observation function of a drift, $f_m(x)$, is an $L^1(\mu)$ function and 
$\int_0^{1/2} f_m(x) d\mu \ne 0$ $(c\ne 0.375)$,
 the distributional limit theorem, Eq. ({\ref{dist.lim.th}), holds. In particular,  the distribution of 
the normalized time average of $f_m(x)$ converges to a Mittag-Leffler distribution:
\begin{equation}
\frac{1}{a_n}\sum_{k=0}^{n-1} f_m \circ T^k \Rightarrow M_{\alpha},
\end{equation}
where the return sequence is given by 
\begin{equation}
 a_n \propto \left\{
 \begin{array}{ll}
   {\displaystyle \frac{n}{\log n}}\quad &(z=2)\\
  \\
   {\displaystyle n^{\alpha}}\quad
&(z>2).
 \end{array}
 \right.
\end{equation}
 Moreover, 
the observation function of the Lyapunov exponent, $g(x) \equiv \ln |T'(x)|= \ln |R'(x)|$, is also the $L^1_+(\mu)$
function. Thus, the normalized Lyapunov exponent is intrinsically random and its distribution converges to 
a Mittag-Leffler distribution. \par
Using numerical simulations, we confirmed the Einstein relation and the constant $\chi_V$, which are in good agreement 
with theory (Fig.~2). Moreover, the relation between the difference of the generalized Lyapunov exponent and the ensemble-averaged 
velocity is also valid except for a large $V$ (Fig.~3). 
The generalized Lyapunov exponent is maximized at $V=0$ and decreased according to the increase in $|V|$.

\begin{figure}
\includegraphics[height=.48\linewidth, angle=0]{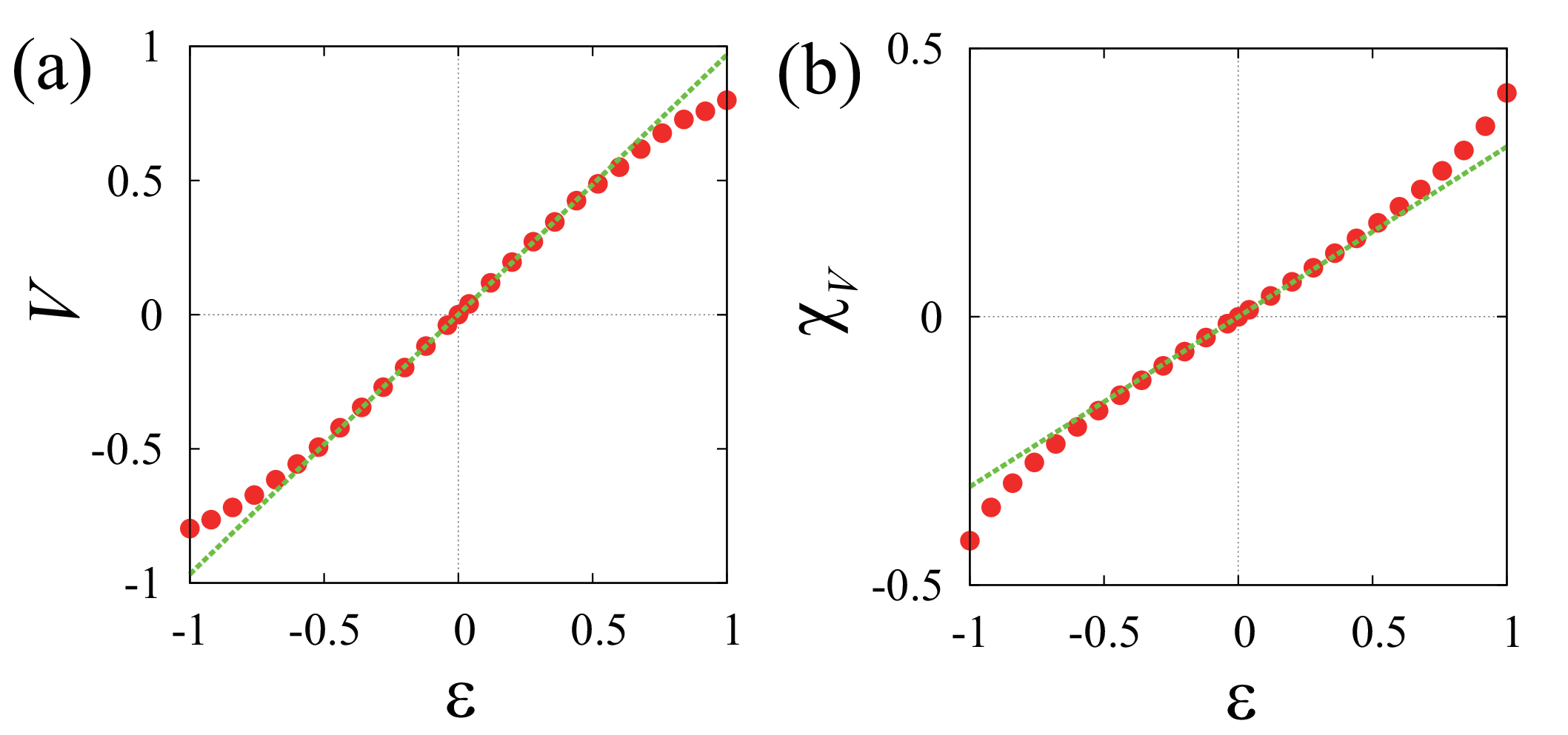}
\caption{ (Color online) Einstein relation and constant $\chi_V$ $(z=2.5)$. Circles show the results of numerical 
simulations. Green lines show the theoretical results obtained by Eq. (\ref{linear.response}) and (\ref{chiV}). Constants $\langle \overline{D}\rangle$ 
and $\langle \overline{\lambda}\rangle$ are obtained by numerical simulations. $K$ is a fitting parameter $(K\cong 1.12)$.}
\end{figure}

\begin{figure}
\includegraphics[height=1.\linewidth, angle=-90]{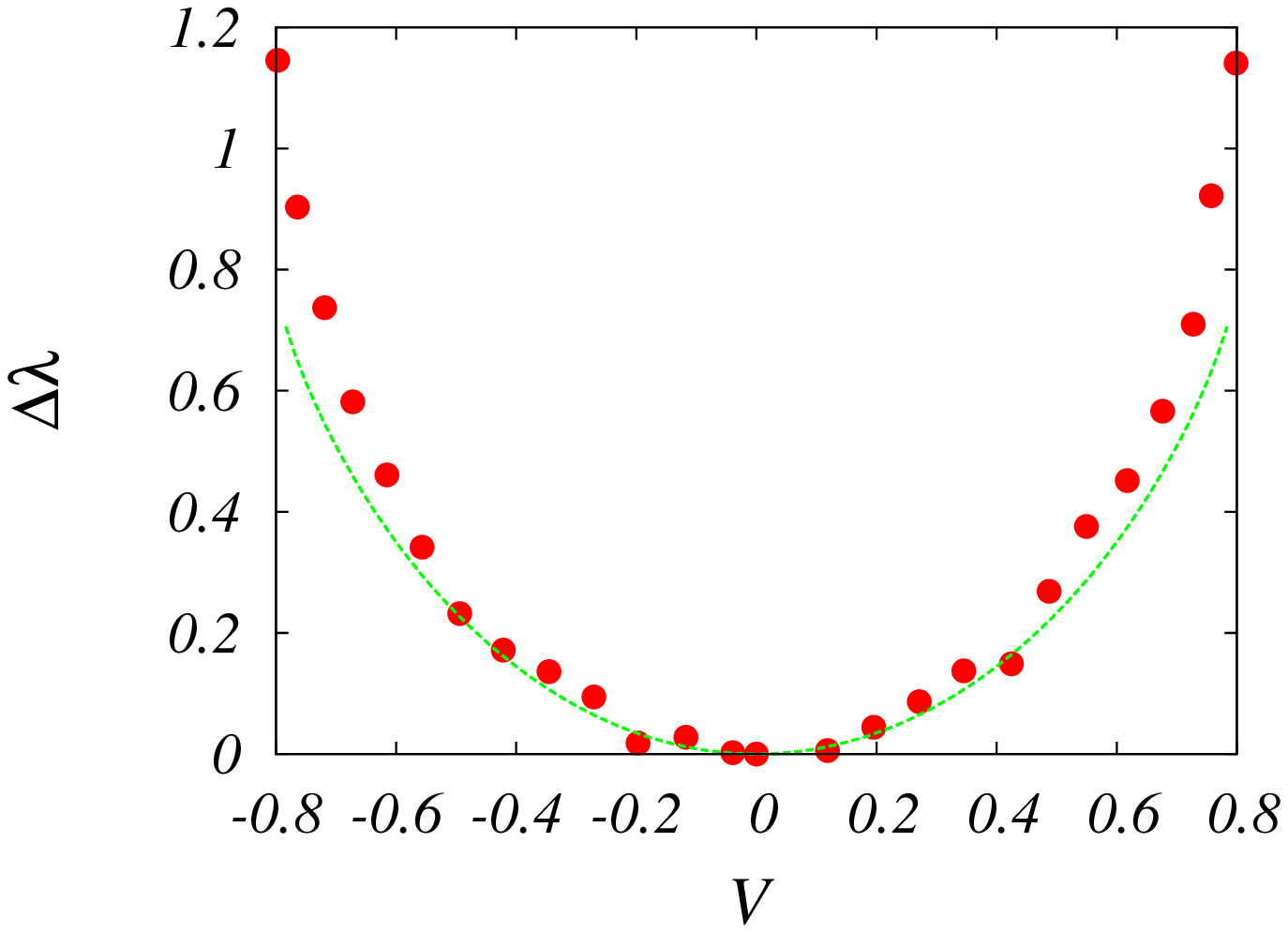}
\caption{ (Color online) Relation between the difference of the generalized Lyapunov exponent and the ensemble-averaged 
velocity $(z=2.5)$. Circles are the results of numerical simulations and the line is the theoretical curve 
(\ref{theoretical.curve}), where $V_{\max}$ and $\mu(J^c)$ are obtained by numerical simulations. In particular, 
we calculate $\mu(J^c)$ by $\lim_{n\rightarrow\infty}\langle \sum_{k=0}^{n} 1_{J^c}(R^kx)/a_n \rangle$,
 and we set $a_n=n^{\alpha}$.}
\end{figure}

{\it Conclusion}.---We derive dynamical systems corresponding to CTRWs.
 In a biased model, we obtain the Einstein relation for the time-averaged
velocity and the time-averaged diffusion coefficient. 
Moreover, the difference of the generalized Lyapunov exponent between unbiased and biased 
dynamical systems is represented by the ensemble-averaged velocity. 
Using Hopf's ergodic theorem, we find that the ratio between the time-averaged velocity and the Lyapunov exponent converges 
to a universal constant. The universal constant is proportional to  bias and the proportional constant is given by the diffusion 
coefficient and the Lyapunov exponent without a bias. 
In general, the ensemble-averaged velocity is represented by the probability $p$: $V=(2p-1)V_{\max}$, and
 $p=\mu(J^c_+)/\mu(J^c)$, where $J^c_+ =[c,1/2]$. 
The relation between $\Delta \lambda$ and $V$ will be universal if the map on cell, $[-1/2+L,L+1/2]$, is continuous
because $\Delta \lambda$ is related to $\mu(J^c_+)/\mu(J^c)$. Moreover, when the derivative of the map on 
the interval representing bias is the same as that of the unbiased map, the constant $K$ is almost unity.
\par

The author thanks T. Miyaguchi for the useful discussions, and Grant-in-Aid for Young Scientists (B) (No. 22740262) for its support.




%

\end{document}